\documentclass[conference]{IEEEtran0}
\usepackage{cite}
\usepackage{graphicx}
\usepackage{url}
\usepackage[centertags]{amsmath}
\usepackage{amssymb,amstext}
\usepackage{amsfonts}
\newcommand{\beq}{\begin{equation}}
\newcommand{\eeq}{\end{equation}}
\newcommand{\beqn}{\begin{align}}
\newcommand{\eeqn}{\end{align}}

\begin{document}

\title{\huge{Per-Antenna Constant Envelope Precoding for Secure Transmission in Large-Scale MISO Systems}}

\author{\IEEEauthorblockN{Jun~Zhu$^{\dag}$, Ning~Wang$^{\S\dag}$, and Vijay~K.~Bhargava$^{\dag}$}
\IEEEauthorblockA{$^{\dag}$Department of Electrical and Computer Engineering,
The University of British Columbia, Vancouver, Canada\\
$^{\S}$School of Information Engineering, Zhengzhou University, Zhengzhou, China
}
}

\IEEEoverridecommandlockouts

\setcounter{page}{1}

\maketitle

\begin{abstract}
Secure transmission in large-scale MISO systems employing artificial noise (AN) is studied
under the per-antenna constant envelope (CE) constraint.
Achievable secrecy rate of the per-antenna CE precoding scheme for large-scale MISO is analyzed
and compared with that of the matched filter linear precoding.
A two-stage per-antenna CE precoding scheme for joint signal-plus-AN transmission is proposed.
The first stage determines the per-antenna CE precoding for the information-bearing signal.
A properly generated AN using an iteration algorithm is incorporated into the transmit signal
in the second stage such that the combined signal-plus-AN satisfies the per-antenna CE constraint
and the AN is orthogonal to the user channel.
It is shown that compared to conventional per-antenna CE transmission,
this joint signal-plus-AN secure transmission scheme does not require additional transmit power.
An alternative low-complexity AN generation scheme which uses a separate antenna to cancel
the AN leakage to the intended user introduced by randomly generated AN is also proposed.
\end{abstract}


\section{Introduction}\label{one}
Multiple-input-multiple-output (MIMO) antenna system is a promising technology
to exploit spatial diversity and improve radio spectral efficiency. 
This is especially important to the design of future high data rate wireless communication systems where
the scarce of spectra 
is becoming a critical limitation.
The MIMO technology has been extensively studied in the literature and has been incorporated into
concurrent wireless communication standards such as LTE and the latest versions of Wi-Fi.
However, as wireless technologies are evolving, small portable devices, e.g. smartphones,
will lead the mobile IP traffic growth \cite{cisco}.
Because it is in general not practical to implement multiple antennas and the associated complex hardware/software
on such small devices due to the size and energy constraints,
advantages of the MIMO technology are greatly forfeited in such scenarios.

Large-scale MIMO, or massive MIMO, is an emerging revolutionary base station (BS) technology
based on MIMO to mitigate the above discrepancies \cite{massive} \cite{massive2}.
It uses a large excess of (typically hundreds of) very low power (in the order of mW) antenna units
at the BS to serve low-complexity single-antenna mobile terminals (MT).
Remarkable improvements in rates as well as in spectral and power efficiency can be achieved
by focusing the radiating power into the ever-smaller MT spots with the very large antenna array.
Massive MIMO is therefore capable of achieving robust performance at low
signal-to-interference-plus-noise ratio (SINR) with highly efficient and inexpensive implementations.

Equipping large antenna array in massive MIMO systems requires each antenna element
and its associated radio-frequency (RF) electronics, e.g. power amplifiers (PAs),
to be inexpensive and power-efficient.
However, cheaply manufactured PAs are in general non-linear devices,
which suffer from linearity issues when processing signals with large amplitude-variations.
A per-antenna constant envelope (CE) nonlinear precoding
was considered in single-user massive MIMO systems in \cite{ce1}.
It was shown that under the per-antenna CE constraint at the BS transmitter,
an equivalent single-input-single-output (SISO) model over additive white Gaussian noise (AWGN) is obtained
for multiple-input-single-output (MISO) system where we have a single-user equipped with a single-antenna \cite{ce1}.
When sufficiently large number of antennas are used,
the corresponding achievable rate under per-antenna CE constraint is close to
the capacity of the MISO channel under the average-only power constraint in high-power regime.
More recently,
the idea of per-antenna CE precoding has been extended to
multi-user massive MIMO systems over flat and frequency-selective fading channels \cite{ce2} \cite{ce3}.

Information security is a critical concern in communication system design.
It is particularly important to wireless communications because of the broadcast nature
of the open media.
The notion of physical-layer (PHY) security has been attracting increasing attention
from both academia and industry.
As a complement to cryptography-based security strategies, the PHY-based approach improves security
of the communication from the {\it perfect secrecy} perspective.
In the case of passive attacks, i.e. eavesdropping, to MIMO systems,
it has been shown that simultaneously transmitting both the information-bearing signal
and properly designed artificial noise (AN) is an efficient way to improve the secrecy rate \cite{an1}.
The application of AN-based secure transmissions for multi-cell massive MIMO systems
has been recently studied in \cite{zhu,zhu2}.
However, design and performance of such security enhancing strategies
under the per-antenna CE constraint have yet been investigated.

In this work, we present our recent investigations of incorporating
AN-based PHY security strategies into large-scale MISO systems under the per-antenna CE constraint.
Different AN generation strategies are considered;
the corresponding improvement to the secrecy capacity is studied.

\section{System Model}\label{two}
\subsection{Constant Envelope Precoding for Large-Scale MISO}\label{2a}
In a MISO system with $N_{t}$ transmit (Tx) antennas and one single receive (Rx) antenna,
the complex channel gain between the $i^{\rm th}$ Tx antenna and the Rx antenna
is $h_{i}$, $i=1,\ldots,N_{t}$.
The total channel vector is $\mathbf{h}=[h_{1},\ldots,h_{N_{t}}]$.
Given the complex transmit signal from the $i^{\rm th}$ antenna be $x_{i}$,
the complex received signal of the user is
\begin{equation}\label{y0}
y = \mathbf{hx} + n = \sum_{i=1}^{N_{t}} h_{i}x_{i} + n,
\end{equation}
where $\mathbf{x}=[x_{1},\ldots,x_{N_{t}}]^{T}$ is the vector form of the Tx signal,
and $n\sim \mathcal{CN}(0,\sigma^{2})$ is a zero-mean circular-symmetric additive white Gaussian noise
(AWGN) at the receiver with variance $\sigma^{2}$.

With conventional linear matched filter (MF) precoder, the Tx signal $\mathbf{x}$ is
the information-bearing signal $s$ multiplied by an MF beamforming vector
$\mathbf{w}\in\mathcal{C}^{N_{t}}$
\begin{equation}\label{xMF}
\mathbf{x}_{MF}=\sqrt{P_{T}}\mathbf{w}s=\sqrt{P_{T}}\frac{\mathbf{h}^{\dag}}{\|\mathbf{h}\|}s,
\end{equation}
where average-only power constraint $P_{T}$ is adopted.
$(\cdot)^{\dag}$ is the conjugate transpose operator,
and $\|\cdot\|$ denotes the 2-norm.
Under the per-antenna CE constraint, as in \cite{ce1},
each antenna transmits at a constant power $\frac{P_{T}}{N_{t}}$.
The Tx signal $x_{i}$ is in the form
$x_{i} = \sqrt{\frac{P_{T}}{N_{t}}} e^{j\theta_{i}^{u}}$, 
where $\theta_{i}^{u}\in(-\pi,\pi]$ is the information-bearing phase
of the $i^{\rm th}$ Tx antenna.
The received signal (\ref{y0}) then becomes
\begin{equation}\label{yce}
y = \sqrt{\frac{P_{T}}{N_{t}}} \sum_{i=1}^{N_{t}} h_{i}e^{j\theta_{i}^{u}} + n.
\end{equation}

When perfect channel state information (CSI) is available at both the transmitter and the receiver,
the received signal $y$ reduces to the output of a SISO system
\begin{equation}\label{yu}
y = \sqrt{P_{T}}\ u + n,
\end{equation}
where $u = \frac{\sum_{i=1}^{N_{t}} h_{i}e^{j\theta_{i}^{u}}}{\sqrt{N_{t}}} $
is the noise-free information signal expected by the intended receiver.
The per-antenna CE precoding maps the information symbol $u$ to
the information-bearing transmit phases $\theta_{i}^{u}$, $i=1,\ldots,N_{t}$,
based on the channel $\mathbf{h}$.
The resulting transmit signal has a constant envelop $\sqrt{\frac{P_{T}}{N_{t}}}$ on every antenna branch.
The objective of per-antenna CE precoding is to determine
$\Theta^{u}=[\theta_{1}^{u},\ldots,\theta_{N_{t}}^{u}]$
given the intended received signal $u$ and the channel $\mathbf{h}$, i.e. the parameterized mapping
\begin{equation}\label{Phi}
\Phi(u;\mathbf{h}) \triangleq \Theta^{u}.
\end{equation}

Given the channel $\mathbf{h}$, all possible noise-free received signal $u$ in (\ref{yu})
of the per-antenna CE transmission lie in the set
\begin{equation}\label{Mh}
\mathcal{M}(\mathbf{h})
\triangleq \left\{\frac{1}{\sqrt{N}}\sum_{i=1}^{N_{t}}h_{i}e^{j\theta_{i}},\
\theta_{i}\in(-\pi,\pi],\ i=1,\ldots,N_{t}\right\}.
\end{equation}
The set $\mathcal{M}(\mathbf{h})$ was shown to have a doughnut-like shape on the complex plain \cite{ce1}.
The key properties of $\mathcal{M}(\mathbf{h})$ include
\begin{enumerate}
\item The set $\mathcal{M}(\mathbf{h})$ is circular symmetric, i.e. $z\in\mathcal{M}(\mathbf{h})$
also implies $ze^{j\varphi}\in\mathcal{M}(\mathbf{h})$ for all $\varphi\in(-\pi,\pi]$.
\item The maximum value of $|u|$ for $u\in\mathcal{M}(\mathbf{h})$ is
\begin{equation}\label{maxMh}
M(\mathbf{h})=\frac{1}{\sqrt{N_{t}}}\sum_{i=1}^{N_{t}}|h_{i}|
=\frac{1}{\sqrt{N_{t}}}\|\mathbf{h}\|_{1}.
\end{equation}
\item The minimum value of $|u|$ for $u\in\mathcal{M}(\mathbf{h})$, denoted by $m(\mathbf{h})$,
is greater than zero and is upper bounded by
\begin{equation}\label{minMh}
m(\mathbf{h})\leq\frac{1}{\sqrt{N_{t}}}\max_{i=1,\ldots,N_{t}}|h_{i}|
=\frac{1}{\sqrt{N_{t}}}\|\mathbf{h}\|_{\infty}.
\end{equation}
\end{enumerate}

\subsection{AN-Based Secure Transmission}\label{2b}
Security of the single-cell single-user MISO downlink is considered in flat-fading.
The BS has size $N_{t}$ antenna array while the MT has only a single antenna.
A single-antenna eavesdropper (Eve) is randomly located in the cell region.
The study can be extended to eavesdropping scenarios where the eavesdropper is equipped with $N_{e}<N_{t}$ antennas.
The eavesdropper aims to retrieve the information transmitted to the mobile terminal.

\begin{figure}
\begin{center}
\includegraphics[width=2.5in, draft=false]{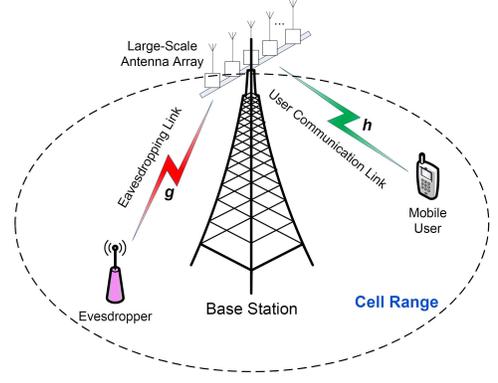}
\caption{The single-user massive MIMO system with a passive eavesdropper.}
\label{ANsystem}
\end{center}
\vskip -8pt
\end{figure}
As a starting point, we first focus on the single-eavesdropper single-antenna case which is shown by Fig. \ref{ANsystem}.
We still use $h_{i}$ to denote the complex channel gain between the $i^{\rm th}$ BS antenna and the MT;
$g_{i}$ is the complex channel gain between the $i^{\rm th}$ BS antenna and the eavesdropper.
${\bf h} \in \mathcal{C}^{N_{t}}$ represents the channel gain row vector between the BS and the MT
with $h_{i}$ as its $i^{\rm th}$ entry.
Similarly, ${\bf g} \in \mathcal{C}^{N_{t}}$ is the channel gain row vector between
the BS and the eavesdropper. 
The BS then can use up to ($N_{t}-1$) degrees of freedom of the antenna array for AN transmission
to degrade Eve's ability of decoding the user message \cite{an1}.

We now illustrate the concept of AN-based secure transmission in conventional linear precoding system
where the average-only power constraint is considered.
The transmitted signal-plus-AN is
\begin{equation}\label{tx_an}
\begin{aligned}
\mathbf{x}
&= \sqrt{\eta P_{T}}\mathbf{w}s + \sqrt{(1-\eta)P_{T}}\mathbf{V}\mathbf{z}\\
&= \sqrt{\eta P_{T}}\mathbf{w}s + \sum_{k=1}^{N_{t}-1}\sqrt{(1-\eta)P_{T}}\mathbf{v}_{k}z_{k}
\end{aligned}
\end{equation}
where
$\eta\in[0,1]$ is the average Tx power allocation factor for user information,
$\mathbf{z}=[z_{1},\ldots,z_{N_{t}-1}]^{T}\in\mathcal{CN}(\mathbf{0}_{N_{t}-1},\mathbf{I}_{N_{t}-1})$
is the AN vector,
and $\mathbf{V} = [\mathbf{v}_{1},\ldots,\mathbf{v}_{N_{t}-1}]\in\mathcal{C}^{N_{t}\times(N_{t}-1)}$
is an AN shaping matrix with $||\mathbf{v}_{k}||=1$, $k=1,\ldots,N_{t}-1$.
The received signal at the mobile terminal is then given by
\begin{equation}\label{yr}
y_{r} = \mathbf{h}(\sqrt{\eta P_{T}}\mathbf{w}s + \sqrt{(1-\eta)P_{T}}\mathbf{V}\mathbf{z}) + n_{r}
\end{equation}
where $n_{r}\sim\mathcal{CN}(0,\sigma_{r}^{2})$ is a circular symmetric complex AWGN.
The eavesdropper's received signal, on the other hand, is
\begin{equation}
y_{e} = \mathbf{g}(\sqrt{\eta P_{T}}\mathbf{w}s + \sqrt{(1-\eta)P_{T}}\mathbf{V}\mathbf{z}) + n_{e}
\end{equation}
where $n_{e}\sim\mathcal{CN}(0,\sigma_{e}^{2})$ is a circular symmetric complex AWGN at the eavesdropper.
The ideal AN shaping matrix $\mathbf{V}$ is chosen such that it lies in the null-space of the user channel $\mathbf{h}$,
i.e. $\mathbf{hV}=\mathbf{0}^{N_{t}-1}$.
Therefore the precoded AN term $\mathbf{Vz}$ will not affect the user reception.
Conversely, an additional noise term is ``seen'' by the eavesdropper,
which degrades her effective signal-to-noise ratio (SNR).
As a consequence, the eavesdropper's capacity is reduced, which results in an improvement of the secrecy capacity
between the BS and the intended user MT.

\section{Security Analysis of Per-Antenna CE Precoding}\label{three}
In this section, we study secrecy capacity of the per-antenna CE precoding and MF precoding.
For MF precoding as given by (\ref{xMF}), the ergodic capacity of the legitimate channel
between the BS and the intended receiving MT is
\begin{equation}\label{C1}
C_{MF} = \mathbb{E}_{\mathbf{h}}
\left[1+\frac{P_{T}}{\sigma^{2}}\|\mathbf{h}\|^{2}\right],
\end{equation}
where the expectation is over the distribution of $\mathbf{h}$.
By assumption, all elements of $\mathbf{h}$ are independent and identically distributed (i.i.d.)
complex Gaussian random variables (RVs) with zero mean and unit variance.
Then $z_{h}=\|\mathbf{h}\|^{2}=\sum_{i=1}^{N_{t}}h_{i}^{2}$ is the sum of
$N_{t}$ i.i.d. exponential RVs with parameter $\lambda=1$,
which follows Gamma distribution with shape parameter $N_{t}$ and scale parameter 1.
The ergodic capacity in (\ref{C1}) is then evaluated as \cite{mckay}
\begin{equation}\label{CMF}
\begin{aligned}
C_{MF}
&= \int_{0}^{\infty}\log_{2}\left(1+\frac{P_{T}}{\sigma^{2}}z\right)\frac{\exp(-z)z^{N_{t}-1}}{\Gamma(N_{t})}dz\\
&= \frac{1}{\ln{2}}\exp\left(\frac{\sigma^{2}}{P_{T}}\right)
   \sum_{n=1}^{N_{t}}E_{n}\left(\frac{\sigma^{2}}{P_{T}}\right),
\end{aligned}
\end{equation}
where $E_{n}(\cdot)$ is the generalized exponential integral \cite{tables}
$E_{n}(x) = \int_{1}^{\infty}\frac{\exp(-xt)}{t^{n}}dt$.
On the other hand, the ergodic capacity of the eavesdropper channel is
\begin{equation}\label{CMFeve}
C_{MF-Eve} = \mathbb{E}_{\mathbf{g,h}}
\left[\log_{2}\left(1+\frac{P_{T}}{\sigma^{2}}
\left|\mathbf{g}\frac{\mathbf{h}^{\dag}}{\|\mathbf{h}\|}\right|^{2}\right)\right].
\end{equation}
Since $\frac{\mathbf{h}^{\dag}}{\|\mathbf{h}\|}$ is a normalized vector with i.i.d. complex Gaussian entries
and it is independent from $\mathbf{g}$, the RV $z_{g}=\mathbf{g}\frac{\mathbf{h}^{\dag}}{\|\mathbf{h}\|}$
has the same distribution as elements of $\mathbf{g}$, i.e. $z_{g}\thicksim \mathcal{CN}(0,1)$.
As a consequence, $|z_{g}|^{2}$ is exponentially distributed with parameter $\lambda=1$,
and the ergodic capacity of the eavesdropper channel is derived as
\begin{equation}
C_{MF-Eve} = \frac{1}{\ln{2}}\exp\left(\frac{\sigma^{2}}{P_{T}}\right)E_{1}\left(\frac{\sigma^{2}}{P_{T}}\right).
\end{equation}
The resulting ergodic secrecy capacity of MF precoding is
\begin{equation}
\label{csecmf}
\begin{aligned}
C_{sec-MF} &= [C_{MF} - C_{MF-Eve}]^{+} \\
&= \left[\frac{1}{\ln{2}}\exp\left(\frac{\sigma^{2}}{P_{T}}\right) \sum_{k=2}^{N_{t}}E_{k}\left(\frac{\sigma^{2}}{P_{T}}\right)\right]^{+},
\end{aligned}
\end{equation}
where $[\cdot]^{+}$ denotes the $\max$ operation $\max\{0,\cdot\}$.

Under the per-antenna CE constraint at the BS transmitter,
the achievable rate of the doughnut channel of the legitimate user, given the channel coefficient $\mathbf{h}$,
is shown to be tightly lower bounded by \cite{ce1}
\begin{equation}\label{CCeh}
\begin{aligned}
C_{CE|\mathbf{h}}
&\geq \log_{2}\left(1+\frac{P_{T}}{\sigma^{2}}
\frac{M(\mathbf{h})-m(\mathbf{h})}{e}\right)\\
&\geq \log_{2}\left(1+\frac{P_{T}}{\sigma^{2}}
\frac{\|\mathbf{h}\|_{1}^{2}-\|\mathbf{h}\|_{\infty}^{2}}{Ne}\right).
\end{aligned}
\end{equation}
By taking expectation of (\ref{CCeh}) over the distribution of the channel $\mathbf{h}$,
we can evaluate the lower bound of the ergodic capacity of the doughnut channel as
\begin{equation}\label{CCe}
C_{CE} \geq \mathbb{E}_{\mathbf{h}}\left[\log_{2}\left(1+\frac{P_{T}}{\sigma^{2}}
\frac{\|\mathbf{h}\|_{1}^{2}-\|\mathbf{h}\|_{\infty}^{2}}{Ne}\right)\right].
\end{equation}
Unfortunately, a closed-form expression for (\ref{CCe}) is difficult to obtain.
Numerical techniques need to be used to evaluate the ergodic capacity lower bound (\ref{CCe}).
Similar to (\ref{yce}), with per-antenna CE precoding, the received signal at the eavesdropper
is given by
\begin{equation}
y_{g} = \sqrt{\frac{P_{T}}{N_{t}}} \sum_{i=1}^{N_{t}} g_{i}e^{j\theta_{i}^{u}} + n.
\end{equation}
Because $\theta_{i}^{u}$, $i=1,\ldots,N_{t}$, are designed according to the mapping (\ref{Phi}),
they are uniformly distributed phases and are independent of the eavesdropper channel $\mathbf{g}$.
The RV $y_{g}$ can thus be written in the form
\begin{equation}
y_{g} = \sqrt{\frac{P_{T}}{N_{t}}} \sum_{i=1}^{N_{t}} g'_{i} + n = \sqrt{P_{T}} \sum_{i=1}^{N_{t}} \frac{g'_{i}}{\sqrt{N_{t}}} + n,
\end{equation}
where $g'_{i}$ have the same distribution as $g_{i}$ because they are obtained by
giving a uniformly distributed random phase rotation to $g_{i}$.
It is straightforward to show that the summation term $g_{s}=\sum_{i=1}^{N_{t}} \frac{g'_{i}}{\sqrt{N_{t}}}$
has the same distribution as $g_{i}$, i.e. zero-mean unit variance complex Gaussian distribution.
Then $|g_{s}|^{2}$ follows exponential distribution with parameter $\lambda=1$.
The ergodic eavesdropper channel capacity for per-antenna CE transmission is therefore
\begin{equation}\label{CeveCe}
\begin{aligned}
C_{CE-Eve}
&= \mathbb{E}_{g_{s}}\left[\log_{2}\left(1+\frac{P_{T}}{\sigma^{2}}|g_{s}|^{2}\right)\right]\\
&= \frac{1}{\ln{2}}\exp\left(\frac{\sigma^{2}}{P_{T}}\right)E_{1}\left(\frac{\sigma^{2}}{P_{T}}\right),
\end{aligned}
\end{equation}
which is identical to the ergodic eavesdropper channel capacity for MF transmission at the BS given by (\ref{CMFeve}).
The ergodic secrecy capacity of the per-antenna CE precoding scheme can then be determined by using
(\ref{CCe}) and (\ref{CeveCe}) as
\begin{equation}
\label{csecce}
C_{sec-CE}=[C_{CE}-C_{CE-Eve}]^{+}.
\end{equation}
It is shown in \cite{ce1} that the per-antenna CE transmission,
by putting more stringent power constraint to the transmit signal,
suffers from a performance loss in capacity compared with matched filtering beamforming
with average-only power constraint.
As $C_{CE-Eve}=C_{MF-Eve}$, the MF precoding will still outperform per-antenna CE precoding
in terms of secrecy capacity of the system.
However, we are going to show that by adding a carefully designed artificial noise,
significant improvement in secrecy capacity can be obtained
without violating the stringent per-antenna CE power constraint.
The benefit of per-antenna CE precoding in less complex RF hardware design is retained.

\section{Per-Antenna CE Precoding for Joint Signal-plus-AN Secure Transmission}\label{four}
\subsection{General Problem Formulation}\label{4a}
\begin{figure}
\begin{center}
\includegraphics[width=1.8in, draft=false]{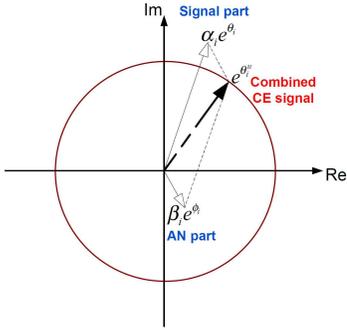}
\caption{The general per-antenna CE transmit signal for secure transmission enabled by AN.}
\label{CEgeneral}
\end{center}
\vskip -8pt
\end{figure}

We adopt the per-antenna CE constraint to the general AN-based secure transmission model in Section \ref{2b}.
The transmit signal of the $i^{\rm th}$ antenna then has the following general form
\begin{equation}\label{CExi}
x_{i} = \sqrt{\frac{P_{T}}{N_t}} \left( \alpha_{i} e^{j\theta_{i}} + \beta_{i} e^{j\phi_{i}} \right)
= \sqrt{\frac{P_{T}}{N_t}} e^{j\theta_{i}^{u}},\
i = 1,\ldots,N_{t},
\end{equation}
where $\alpha_{i}$ and $\theta_{i}$ are the amplitude and phase of the user information part of
the per-antenna CE signal on the $i^{\rm th}$ antenna,
$\beta_{i}$ and $\phi_{i}$ are the amplitude and phase of the corresponding
AN part transmitted by the same antenna.
$\theta_{i}$, $\phi_{i}$, and $\theta_{i}^{u}$ are phase angles within the range $(-\pi,\pi]$.
In order to guarantee the second equality in (\ref{CExi}) such that the per-antenna CE constraint holds,
we must have the following trigonometric relation
\begin{equation}\label{trig}
\alpha_{i}^{2}+\beta_{i}^{2}+2\alpha_{i}\beta_{i}\cos(\theta_{i}-\phi_{i})=1
\end{equation}
for all $i=1,\ldots,N_{t}$.
The relationship (\ref{trig}) is illustrated by Fig. \ref{CEgeneral},
where the unit circle represents the normalized per-antenna CE signal envelope.

Ideally the AN part of the transmit signal (\ref{CExi}) is ``invisible'' to the user, which requires
\begin{equation}\label{AN0}
\sqrt{\frac{P_{T}}{N_t}}\sum_{i=1}^{N_{t}}h_{i}\beta_{i}e^{j\phi_{i}}=0.
\end{equation}
On the other hand, the noise-free user information part of the received signal, denoted
\begin{equation}\label{uce}
u = \sqrt{\frac{1}{N_{t}}} \sum_{i=1}^{N_{t}} \alpha_{i}h_{i}e^{j\theta_{i}},
\end{equation}
is in the alphabet $\mathcal{U}$ of the information-bearing symbol expected by the receiver.
In addition, given the power allocation factor $\eta$ for joint signal and AN transmission,
the parameters $\alpha_{i}$ and $\beta_{i}$ should also satisfy the equality condition
\begin{equation}\label{pallc}
\frac{\sum_{i=1}^{N_{t}}\alpha_{i}^{2}}{\sum_{i=1}^{N_{t}}\beta_{i}^{2}} = \frac{\eta}{1-\eta}.
\end{equation}
The general problem for joint signal-plus-AN secure transmission in such a per-antenna CE system is to determine
the parameters ($\alpha_{i},\theta_{i}$) and ($\beta_{i},\phi_{i}$) for all $N_{t}$ antenna branches such that
the secrecy capacity of the system with respect to the eavesdropper as in Section \ref{2b} is maximized.

However, because the parameter pairs ($\alpha_{i},\theta_{i}$) and ($\beta_{i},\phi_{i}$) are coupled through
the nonlinear CE constraint characterized by the relationships (\ref{trig}) and (\ref{pallc}),
the problem cannot easily be transformed into a well-formulated per-antenna CE precoding problem as in \cite{ce1}.
Some techniques must be employed to decouple the information-bearing signal and the artificial noise
while achieving the design objective.
We will address this problem in our proposed design.

\subsection{Two-Stage Per-Antenna CE Precoding for Joint Signal-plus-AN Transmission}\label{4b}
A straightforward way to decouple the information-bearing signal and the artificial noise is to design one part first
based on certain criteria and then determine the other accordingly such that the CE constraint holds.
In a communication system we first need to focus on the user information which should meet certain requirements.
Conversely requirement for the AN is usually loose.
We only try to introduce as much extra noise to the eavesdropper as possible while
minimizing the AN's impact on user information reception.
The aggregate AN at the receiver should sum to zero such that (\ref{AN0}) holds.
Therefore we propose to determine the user signal first,
which is then used to determine the AN to be transmitted according to the per-antenna CE constraint.

As illustrated in Section \ref{2b}, ideally the AN shaping matrix $\mathbf{V}$ should be in the null-space
of the user channel such that the aggregate AN at the receiver is equal to zero,
regardless the random noise $\mathbf{z}$.
However, this constraint is too stringent for AN-based secure transmission in single-user large-scale MISO systems.
Note that the generated AN is deterministic for each transmission.
Therefore the null-space requirement is not necessary.
All we need is (\ref{AN0}), which requires the aggregate AN at the receiver is zero.

We consider per-antenna CE precoding for the user information as in \cite{ce1},
i.e. we set $\alpha_{i}=1$ for all $i=1,\ldots,N_{t}$.
The precoder mapping $\Phi(u)=\Theta^{u}$ proposed in \cite{ce1} is employed to find $\theta_{i}$'s.
It was illustrated in \cite{ce1} that with large number of transmit antennas, which is the case of massive MIMO,
there are low-complexity gradient-based algorithms which gives $\Theta^{u}$ with small error norms
$\left|u-\sum_{i=1}^{N_{t}}h_{i}e^{j\theta_{i}}\sqrt{N_{t}}\right|^{2}$.

With $\alpha_i=1,\forall\ i$, 
(\ref{trig}) is simplified to
\begin{equation}
\beta^2_i+2\beta_i \cos(\theta_i-\phi_i)=0 \stackrel{\beta_i \neq 0}{\longrightarrow} \beta_i=-2 \cos(\theta_i-\phi_i).
\end{equation}
Accordingly, the ``invisible'' condition (\ref{AN0}) is rewritten as (by ignoring the scaling factors)
\begin{equation}\label{AN02}
\sum_{i=1}^{N_{t}}h_{i}\cos(\theta_i-\phi_i)e^{j\phi_{i}}=0,
\end{equation}
in which $\phi_i,i=1,\ldots,N_t$ is the only variable, given that $\theta_i,i=1,\ldots,N_t$ have already been found
based on the precoder mapping $\Phi(u)=\Theta^{u}$. 
With sufficiently large $N_t$, we can always find a set of $\phi_i,i=1,\ldots,N_t$
satisfying the ``invisible'' condition given in (\ref{AN0}).
In this case, we obtain a ``free'' secrecy improvement from the power consumption perspective,
as the AN is designed to meet the trigonometric relationship in (\ref{trig}).

\subsection{Random-Phase AN Generation and AN Leakage Cancellation}\label{4c}
Design of an optimal CE precoding mapping for the AN generation scheme proposed in Section \ref{4b},
from the secrecy rate perspective, is nontrivial.
Complexity of the algorithm therefore may become a big issue in practical implementations.
In this subsection, we propose an alternative scheme with low complexity, 
which is enabled by one single additional antenna element employing more sophisticated hardware.

\begin{figure}
\begin{center}
\includegraphics[width=1.8in, draft=false]{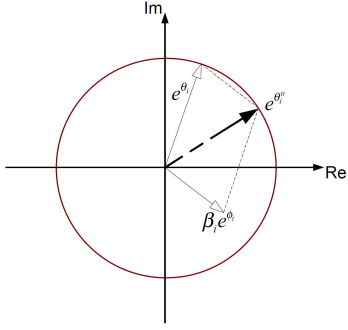}
\caption{The random-phase AN generation and amplitude adjustment for per-antenna CE transmission.}
\label{randomAN}
\end{center}
\vskip -8pt
\end{figure}

A uniform AN phase $\phi_{i}$ is generated after determining the signal phase $\theta_{i}$.
The AN amplitude which preserves the per-antenna CE property of the combined transmit symbol is calculated accordingly.
This is named the random-phase AN scheme, and the general idea of random-phase AN generation
and the amplitude adjustment for per-antenna CE transmission is shown in Fig. \ref{randomAN}.

As in the joint signal-plus-AN per-antenna CE secure transmission scheme in Section \ref{4b},
the signal part is designed by adopting the per-antenna CE precoding.
The AN phase $\phi_{i}$ is randomly generated and uniformly distributed on $(-\pi,\pi]$.
The AN amplitude is determined by the phase difference between the per-antenna CE signal and the random AN phase,
as illustrated in Fig. \ref{randomAN}.
The aggregated AN at the intended receiver, given by
\begin{equation}\label{ANleakage1}
y_{rAN1} = \sqrt{\frac{P_{T}}{N_t}}\sum_{i=1}^{N_{t}}h_{i}\beta_{i}e^{j\phi_{i}},
\end{equation}
is most likely non-zero.
AN leakage will thus occur, which degrades the secrecy capacity.  
Note that (\ref{ANleakage1}) is simply a scalar.
Therefore we can always add one single transmit antenna with user channel gain $h_{0}$ and transmit an additional AN signal
\begin{equation}
x_{0} = \frac{-y_{rAN}}{h_{0}} = \beta_{0}e^{j\phi_{0}}
\end{equation}
such that the overall aggregate AN at the receiver is equal to zero
\begin{equation}\label{ANcancel}
\sqrt{\frac{P_{T}}{N_t}}\sum_{i=0}^{N_{t}}h_{i}\beta_{i}e^{j\phi_{i}} = h_{0}x_{0}+y_{rAN} = 0.
\end{equation}
That is, we need only one extra antenna which does not satisfy the per-antenna CE constraint in the system
to completely cancel the AN leakage due to the random-phase AN and CE constraint on the other $N_{t}$ antennas.

\subsection{Secrecy Rate Evaluation}\label{4d}
By assuming per-antenna CE information signals,
the legitimate user's rate expression is the same as that in \cite{ce1}.
Therefore, when calculating the achievable secrecy rate in this work,
the key is in how to evaluate the eavesdropper capacity.
We next show an upper bound of the eavesdropper capacity by taking the correlation
between two RVs, i.e. the information power and the AN power into account.
This upper bound is non-trivial for the AN cancellation scheme given in Section \ref{4c} (denoted Scheme II),
but becomes trivial for the scheme of Section \ref{4b} (denoted Scheme I).
More effective ways to evaluate the eavesdropper capacity under the nonlinear per-antenna CE constraint,
which remains an open problem, will be investigated in the journal version.

For both AN generation schemes,
the eavesdropper's received signal can be expressed as
\begin{equation}\label{yeve}
\begin{aligned}
y_{\rm eve}
&=\sqrt{\frac{P_{T}}{N_t}}\sum^{N_t}_{i=1} g_i e^{j \theta_i}+\sqrt{\frac{P_{T}}{N_t}}\sum_{i=0}^{N_t} \beta_i g_i e^{j \phi_i}+n \\
&=\sqrt{P_{T}}u_{\rm eve}+n_{\rm AN}+n,
\end{aligned}
\end{equation}
where we define $u_{\rm eve}=\sqrt{\frac{1}{N_t}}\sum^{N_t}_{i=1} g_i e^{j \theta_i}$ and
$n_{\rm AN}=\sqrt{\frac{P_{T}}{N_t}}\sum_{i=0}^{N_t} \beta_i g_i e^{j \phi_i}$.
The term $n$ is the AWGN, which has zero mean and variance $\sigma^2$.
For Scheme I, we have $\beta_0=0$.

The above eavesdropper channel model (\ref{yeve}) is similar to the equivalent SISO channel model for the legitimate MISO system demonstrated in \cite{ce1}.
The eavesdropper capacity is thus given by
\begin{equation}
C_{CE,E}=\underset{u}{\mbox{sup}}\ I(u;u_{\rm eve}),
\end{equation}
where the function $I(\cdot;\cdot)$ gives the mutual information between two RVs.
According to the eavesdropper's equivalent SISO model (\ref{yeve}), we have the following Markov relationship
\begin{equation}\label{markov}
u \leftrightarrow \Theta \leftrightarrow y_{\rm eve} \leftrightarrow u_{\rm eve}.
\end{equation}
By the data processing inequality \cite{coverbook}, we must have $I(u;u_{\rm eve}) \leq I(\Theta;y_{\rm eve})$.
Therefore we have the following upper bound of the eavesdropper capacity
\begin{equation}
\label{CCEE}
\begin{aligned}
C_{CE,E}&=\underset{u}{\mbox{sup}}\ I(u;u_{\rm eve}) \\
&\leq \underset{\Theta}{\mbox{sup}}\ I(\Theta;y_{\rm eve})
\leq \log_2 \left(1+\frac{P_{u_{\rm eve}}}{P_{n_{\rm AN}}+\sigma^2}\right),
\end{aligned}
\end{equation}
where $P_{n_{\rm AN}}=\mathbb{E}_{\Theta,\Phi}\bigg[\left(u_{\rm eve}+n_{\rm AN}\right)^2\bigg]-\mathbb{E}_{\Theta}\bigg[u_{\rm eve}^2\bigg]$ and $P_{u_{\rm eve}}=\mathbb{E}_{\Theta}\bigg[u_{\rm eve}^2\bigg]$.
Note that the above power splitting approach
overestimates the information signal power, which results in an upper bound on Eve capacity.
By redefining the power terms in (\ref{CCEE}) as $P_{u_{\rm eve}}=\mathbb{E}_{\Theta,\Phi}\bigg[\left(u_{\rm eve}+n_{\rm AN}\right)^2\bigg]-\mathbb{E}_{\Theta}\bigg[n_{\rm AN}^2\bigg]$ and $P_{n_{\rm AN}}=\mathbb{E}_{\Theta}\bigg[n_{\rm AN}^2\bigg]$, a lower bound on the eavesdropper capacity can be obtained instead.

It can be shown that based on the above eavesdropper capacity upper bound and the corresponding signal-AN power splitting,
the approximate power terms for Scheme I are given as
\begin{equation}\label{P1}
P_{u_{\rm eve}}=P_T \|{\bf g}\|^2/N_t,\quad P_{n_{\rm AN}}=0,
\end{equation}
which completely ignores the AN power and therefore reduces to a trivial upper bound.
By substituting (\ref{P1}) into (\ref{CCEE}), the capacity expression is identical to
that for per-antenna CE precoding (without AN), given in Section \ref{three}.

The power terms for the eavesdropper capacity upper bound with Scheme II can be calculated in a similar way.
\begin{eqnarray}\label{P2}
P_{u_{\rm eve}} &&\hskip-18pt = P_T \|{\bf g}\|^2/N_t,\\
P_{n_{\rm AN}} &&\hskip-18pt = \frac{P_T}{N_t} \cdot \mathbb{E}_{\Theta,\Phi}\left[\left(g_0\cdot \frac{\sum_{i=1}^{N_t} h_i \beta_i e^{j \theta_i}}{h_0}\right)^2\right].
\end{eqnarray}
In both schemes, $u_{\rm eve}$ are zero-mean Gaussian distributed with variance $P_{u_{\rm eve}}$.
As $N_t \to \infty$, according to the Central Limit Theorem (CLT), we have
\begin{equation}
\sqrt{\frac{P_T}{N_t}}\cdot \frac{g_0\sum_{i=1}^{N_t} h_i \beta_i e^{j \theta_i}}{h_0}
\overset{N_{t}\rightarrow\infty}{\longrightarrow} \mathcal N(0,P_{n_{\rm AN}}).
\end{equation}
The resulting eavesdropper capacity upper bound for Scheme II is thus non-trivial.

\section{Numerical Results}\label{five}
\begin{figure}
\begin{center}
\includegraphics[width=3.5in, draft=false]{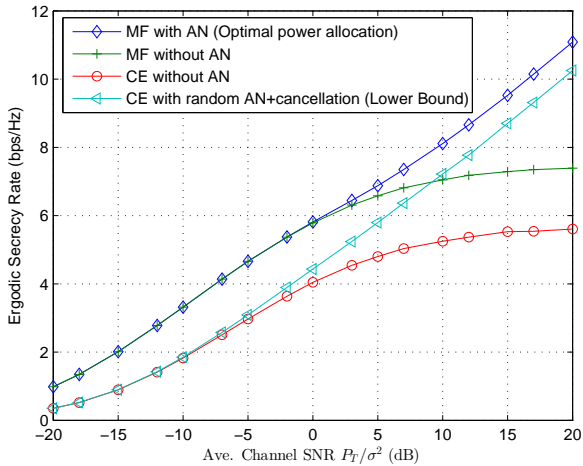}
\caption{Ergodic secrecy capacity v.s. $P_T/\sigma^2$ for i.i.d. Rayleigh fading with $N_t=100$.}
\label{fig1}
\end{center}
\vskip -8pt
\end{figure}

In this section, we evaluate the performance of the proposed secure constant envelope large-scale MISO system
in terms of ergodic secrecy capacity as a function of average channel SNR, $\frac{P_T}{\sigma^2}$.
In Fig.~\ref{fig1}, the curves achieved by MF precoding (MF without AN in Fig.~\ref{fig1}) and per-antenna CE precoding
(CE without AN in Fig.~\ref{fig1}) are calculated based on (\ref{csecmf}) and (\ref{csecce}), respectively.
For the curves achieved by AN-based schemes, including MF precoding with AN and optimal power allocation for the maximization of secrecy capacity (MF with AN in Fig.~\ref{fig1}), and Scheme II (CE with random AN plus cancellation, discussed in Section IV-C) are obtained by Monte-Carlo simulations.

According to Fig.~\ref{fig1}, we first observe that the performance for linear MF
and per-antenna CE transmission (without AN) saturate at high-$\frac{P_T}{\sigma^2}$ regime,
which motivates the AN-based transmission scheme for security improvement.
In contrast,
%
Scheme II described in Section \ref{4c} results in a significantly higher secrecy capacity,
especially in the high-$\frac{P_T}{\sigma^2}$ regime.
The antenna element for leakage cancellation serves as an additional source of AN to the eavesdropper
to mask the legitimate communication.
The proposed method leads to comparable secrecy capacity performance with conventional MF precoded transmission
with AN (optimal power allocation between data and AN \cite{mckay}).
We consider it a non-trivial observation, as with the absence of AN, the secrecy capacity gap between MF precoding
and per-antenna CE precoding is fairly large, especially in the high-$\frac{P_T}{\sigma^2}$ regime.
It indicates that with the proposed AN generation scheme,
the cheaper and more power efficient CE implementation of large-scale MIMO
can achieve secure transmission with performance very close to linear MF precoder with AN.
The only cost is one extra antenna violating the per-antenna CE constraint.
Although not shown in our preliminary work, Scheme I of Section \ref{4b}
is expected to achieve a similar performance as Scheme II,
with an expense of $N_t$ iterations to find a group of $\phi_i$'s satisfying (\ref{AN0}).

\section{Conclusions}\label{six}
In this work, we have investigated secure transmission in large-scale MISO systems
under the per-antenna constant envelope constraint.
Based on a security analysis of the per-antenna CE precoding scheme for large-scale MISO,
we have proposed to use joint signal-plus-artificial-noise transmission to improve security
in systems under the per-antenna CE constraint.
A two-stage per-antenna CE precoding framework for joint signal-plus-AN transmission was proposed.
In the first stage we determine the per-antenna CE precoding for the information-bearing signal.
A properly generated artificial noise is incorporated into the transmit signal such that
the combined signal-plus-AN satisfies the CE constraint and the AN part is ``invisible'' to the legitimate user.
It is shown that compared to conventional per-antenna CE transmission,
the joint signal-plus-AN secure transmission schemes do not require additional transmit power.
A low-complexity AN generation scheme which requires an additional antenna element to cancel
the AN leakage to the intended user introduced by randomly generated AN was also investigated.



\end{document}